\documentclass[useAMS,usenatbib]{mn2e}
\pdfoutput=1
\usepackage[pdftex,pdfpagemode={UseOutlines},bookmarks,bookmarksopen,colorlinks,linkcolor={blue},citecolor={green},urlcolor={red}]{hyperref}
\usepackage{journals_macros}
\usepackage{graphicx}
\usepackage{times}
\usepackage{xspace}

\newcommand{\hmpc}{$\,{\rm h}^{-1}$ Mpc\xspace}
\newcommand{\wprp}{$w_p(r_p)$\xspace}

\title[Two-point correlations in the COSMOS field]{Understanding the shape of the
  galaxy two-point correlation function at $z\simeq1$ in the COSMOS field}

\author[S. de la Torre et al.]
 {S. de la Torre,$^{2,3}$\thanks{E-mail: sylvain.delatorre@brera.inaf.it}
  L. Guzzo,$^{2}$
  K. Kova{\v c},$^{8}$
  C. Porciani,$^{4}$
  U. Abbas,$^{7}$
  B. Meneux,$^{5,6}$
  \newauthor
  C.M. Carollo,$^{8}$
  T. Contini,$^{9}$
  J.-P. Kneib,$^{1}$
  O. Le F\`evre,$^{1}$
  S.J. Lilly,$^{8}$
  V. Mainieri,$^{10}$
  \newauthor
  A. Renzini,$^{11}$
  D. Sanders,$^{21}$
  M. Scodeggio,$^{3}$
  N. Scoville,$^{22}$
  G. Zamorani,$^{12}$
  S. Bardelli,$^{12}$
  \newauthor
  M. Bolzonella,$^{12}$
  A. Bongiorno,$^{5}$
  K. Caputi,$^{8}$
  G. Coppa,$^{12}$
  O. Cucciati,$^{1}$
  L. de Ravel,$^{1}$
  \newauthor
  P. Franzetti,$^{3}$
  B. Garilli,$^{3}$
  A. Iovino,$^{2}$
  P. Kampczyk,$^{8}$
  C. Knobel,$^{8}$
  A.M. Koekemoer,$^{13}$
  \newauthor
  F. Lamareille,$^{9}$
  J.-F. Le Borgne,$^{9}$
  V. Le Brun,$^{1}$
  C. Maier,$^{8}$
  M. Mignoli,$^{12}$
  R. Pell\'o,$^{9}$
  \newauthor
  Y. Peng,$^{8}$
  E. Perez-Montero,$^{9}$ 
  E. Ricciardelli,$^{14}$
  J. Silverman,$^{8}$
  M. Tanaka,$^{10}$
  L. Tasca,$^{3}$
  \newauthor
  L. Tresse,$^{1}$
  D. Vergani,$^{12}$
  N. Welikala,$^{1}$
  E. Zucca,$^{12}$
  D. Bottini,$^{3}$
  A. Cappi,$^{12}$
  \newauthor
  P. Cassata,$^{15}$
  A. Cimatti,$^{16}$
  M. Fumana,$^{3}$
  O. Ilbert,$^{1}$
  A. Leauthaud,$^{17}$
  D. Maccagni,$^{3}$
  \newauthor
  C. Marinoni,$^{18}$
  H.J. McCracken,$^{19}$
  P. Memeo,$^{3} $
  P. Nair,$^{12}$
  P. Oesch,$^{8}$
  L. Pozzetti,$^{12}$
  \newauthor
  V. Presotto,$^{2}$
  and R. Scaramella$^{20}$ \\
  $^{1}$ Laboratoire d'Astrophysique de Marseille, Marseille, France \\
  $^{2}$ INAF - Osservatorio Astronomico di Brera, Milano, Italy \\
  $^{3}$ INAF - Istituto di Astrofisica Spaziale e Fisica Cosmica di Milano,
  Milano, Italy \\
  $^{4}$ Argelander Institute for Astronomy, University of Bonn, Bonn, Germany \\
  $^{5}$ Max Planck Institut f\"ur Extraterrestrische Physik, Garching,
  Germany \\
  $^{6}$ Universitats-Sternwarte, Muenchen, Germany \\
  $^{7}$ INAF - Osservatorio Astronomico di Torino, Pino Torinese, Italy \\
  $^{8}$ Institute of Astronomy, ETH Zurich, Zurich, Switzerland \\
  $^{9}$ Laboratoire d'Astrophysique de l'Observatoire Midi-Pyr\'en\'ees,
  Toulouse, France \\
  $^{10}$ European Southern Observatory, Garching, Germany \\
  $^{11}$ INAF - Osservatorio Astronomico di Padova, Padova, Italy \\
  $^{12}$ INAF - Osservatorio Astronomico di Bologna, Bologna, Italy \\
  $^{13}$ Space Telescope Science Institute, Baltimore, USA \\
  $^{14}$ Dipartimento di Astronomia, Universit\'a di Padova, Padova, Italy \\
  $^{15}$ Department of Astronomy, University of Massachusetts, Amherst, USA \\
  $^{16}$ Dipartimento di Astronomia, Universit\'a di Bologna, Bologna, Italy \\
  $^{17}$ Berkeley Lab \& Berkeley Center for Cosmological Physics, University
  of California, Berkeley, USA \\
  $^{18}$ Centre de Physique Th\'eorique de Marseille, Marseille, France \\
  $^{19}$ Institut d'Astrophysique de Paris, Paris, France \\
  $^{20}$ INAF - Osservatorio Astronomico di Roma, Roma, Italy \\
  $^{21}$ Institute for Astronomy, University of Hawaii, Honolulu, USA \\
  $^{22}$ Astronomy Department, Caltech, Pasadena, USA}

\begin{document}

\date{Accepted 2010 July 12.  Received 2010 July 7; in original form 2010 May 21}

\pagerange{\pageref{firstpage}--\pageref{lastpage}} \pubyear{2010}

\maketitle

\label{firstpage}

\begin{abstract}

We investigate how the shape of the galaxy two-point correlation function as
measured in the zCOSMOS survey depends on local environment, quantified in
terms of the density contrast on scales of 5\hmpc.  We show that the flat
shape previously observed at redshifts between $z=0.6$ and $z=1$ can be
explained by this volume being simply 10\% over-abundant in high-density
environments, with respect to a Universal density probability distribution
function.  When galaxies corresponding to the top 10\% tail of the
distribution are excluded, the measured \wprp steepens and becomes
indistinguishable from $\Lambda CDM$ predictions on all scales.  This is the
same effect recognised by Abbas \& Sheth in the SDSS data at $z\simeq 0$ and
explained as a natural consequence of halo-environment correlations in a
hierarchical scenario.  Galaxies living in high-density regions trace dark
matter halos with typically higher masses, which are more correlated.  If the
density probability distribution function of the sample is particularly rich
in high-density regions because of the variance introduced by its finite size,
this produces a distorted two-point correlation function. We argue that this
is the dominant effect responsible for the observed ``peculiar'' clustering in
the COSMOS field.

\end{abstract}


\section{Introduction}

Advances in the spectroscopic survey capabilities of 8-meter class telescopes
have allowed us in the recent years to extend detailed studies of the
clustering of galaxies to the $z\simeq1$ Universe
\citep{coil04,lefevre05b,pollo06,coil06,meneux06,delatorre07,meneux08,coil08,abbas10}.
The most recent contribution to this endeavour is the COSMOS survey
\citep{scoville07}, and in particular zCOSMOS, its redshift follow-up with
VIMOS at the ESO-VLT \citep{lilly07}.

Early angular studies of the COSMOS field \citep{mccracken07} and more recent
analyses of the first ten thousand zCOSMOS redshifts to $I_{AB}=22.5$, have
evidenced significant ``excess'' clustering in the large-scale shape of the
two-point angular and projected correlation function. The redshift
  information from zCOSMOS, in particular, shows this excess to dominate in
the redshift range $0.5<z<1$ \citep{meneux09}.  More precisely, the shape of
the projected two-point correlation function \wprp appears to decay much less
rapidly than observed at similar redshifts in independent data as the
VVDS survey \citep{meneux08} and with respect to predictions of standard
$\Lambda$CDM cosmology as incarnated by the Millennium simulation
\citep{delucia07,kitzbichler07}. The observed flat shape\footnote{
  $\gamma\sim 1.5$ instead of the $\gamma\sim 1.8$ expected when approximating
  $\xi(r)$ with a power law (i.e. $\xi(r)=\left(r_0/r\right)^\gamma$) below
  $r=10$\hmpc} is difficult to reconcile with the theory, unless an
unrealistic scale-dependent bias between galaxies and matter is advocated.
While plausibly related to the presence of particularly rich large-scale
structures dominating the COSMOS volume around $z\simeq 0.7$
\citep[e.g.][]{meneux09,guzzo07}, this effect still awaits a quantitative
explanation.

In a recent series of papers, \citet{abbas05,abbas06,abbas07} have used the
Sloan Digital Sky Survey \citep[SDSS,][]{york00}, together with {\it Halo
  Occupation Distribution} models \citep[HOD, e.g.][]{cooray02} to show how in
general the amplitude and shape of the galaxy correlation function depend on
the {\sl environment} in which the galaxies are found.  Once a local density
is suitably defined over a given scale, galaxies living in over-dense regions
show a stronger clustering than those in average or under-dense environments.
This is shown to be a consequence of the direct correlation arising in
hierarchical clustering between the mass of the dark matter halos in which
galaxies are embedded, and their large-scale environment: the mass function of
dark-matter halos is top-heavy in high-density regions, thus selecting
galaxies in these environments we are selecting halos of higher mass, which
are more clustered.  The net result is to introduce a \emph{scale-dependent
  bias} in the observed correlation function, when this is compared to the
expected dark-matter clustering \citep{abbas06,abbas07}.

In this Letter we investigate whether this effect is at work also at
$z\simeq0.7$ and could explain quantitatively the observed shape of \wprp in
the zCOSMOS data.

\section{Data and Methods} \label{sec:data}

\subsection{The zCOSMOS 10k-bright sample}

zCOSMOS is a large spectroscopic survey performed with the VIsible
Multi-Object Spectrograph \citep[VIMOS,][]{lefevre03} at the ESO--VLT. The
zCOSMOS-bright survey \citep{lilly07} has been designed to follow-up
spectroscopically the entire $1.7~deg^2$ COSMOS-ACS field
\citep{scoville07,Koekemoer07} down to $I_{AB}=22.5$. We use in this analysis
the first-epoch set of redshifts, usually referred to as the \emph{zCOSMOS
  10k-bright sample} (``10k sample'', hereafter), including $10,644$
galaxies. At this magnitude limit, the survey redshift distribution peaks at
$z\simeq0.6$, with a tail out to $z\simeq 1.2$. We only consider secure
redshifts, i.e. confidence classes $4.\mathsf{x}$, $3.\mathsf{x}$, $9.3$,
$9.5$, $2.4$, $2.5$, and $1.5$, representing $88\%$ of the full 10k sample
\citep[see][for details]{lilly09} and 20.4\% of the complete
  $I_{AB}<22.5$ magnitude-limited parent sample over the same area. These
data are publicly available through the ESO Science Data Archive
site\footnote{http://archive.eso.org/cms/eso-data/data-packages/}.

\subsection{Mock galaxy surveys}

In addition to the observed data, in this analysis we also make use of a set
of 24 mock realisations of the zCOSMOS survey, constructed combining the
Millennium Run N-body
simulation\footnote{http://www.mpa-garching.mpg.de/millennium/}, with a
semi-analytical recipe of galaxy formation \citep{delucia07}.  The Millennium
Run is a large dark matter N-body simulation that follows the hierarchical
evolution of $2160^3$ particles between $z=127$ and $z=0$ in a cubic volume of
$500^3~h^{-3}~Mpc^3$. It assumes a concordance cosmological $\Lambda CDM$
model with $(\Omega_m,~\Omega_\Lambda,~\Omega_b,~h,~n,~\sigma_8) =
(0.25,~0.75,~0.045, ~0.73,~1,~0.9)$. The resolution of the N-body simulation,
$8.6 \times 10^8~h^{-1} M_\odot$, coupled with the semi-analytical model
allows one to resolve with a minimum of 100 particles halos containing
galaxies with a luminosity of $0.1L^*$ \citep[see][]{springel05}.  Galaxies
are generated inside these dark matter halos using the semi-analytic model of
\citet{croton06}, as improved by \citet{delucia07}. This model includes the
physical processes and requirements originally introduced by \citet{White91}
and refined by \citet{kauffmann00}, \citet{springel01}, \citet{delucia04}, and
\citet{springel05}.  Twenty-four mocks are created, and then ``observed'' as
to reproduce the zCOSMOS selection function \citep[][]{iovino10}.

\subsection{Local density estimator}

To characterise galaxy environment we use the dimensionless density contrast
measured by \citet{kovac10} around each galaxy in the sample. For each galaxy
at a comoving position ${\bf r}$ we compute the dimensionless 3D density
contrast smoothed on a scale $R$, $\delta_g({\bf r},R)=(\rho({\bf
  r},R)-\bar{\rho}({\bf r}) )/\bar{\rho}({\bf r})$, where $\rho({\bf r},R)$ is
the density of galaxies measured on a scale $R$ and $\bar{\rho}({\bf r})$ is
the overall mean density at ${\bf r}$.  $\rho({\bf r},R)$ is estimated around
each galaxy of the sample by counting objects within an aperture (defined
either through a top-hat of size $R$ or a Gaussian filter with similar
dispersion). The reconstructed over-densities are properly corrected for the
survey selection function and edge effects. \citet{kovac10} studied different
density estimators, corresponding to varying galaxy tracers, filter shapes and
smoothing scales.  Here we use $\delta_g$ as reconstructed with a Gaussian
filter with dispersion $R=5$\hmpc. Note that the mass enclosed by such filter
is equal to that inside a top-hat filter of size $\sim7.8$\hmpc. We refer the
reader to \citet{kovac10} for a full description of the technique.

\subsection{Expected probability distribution function of the density contrast}

The density contrast distribution can be predicted analytically using some
approximations. Empirically, it has been found that the probability
distribution function (PDF) of the mass density contrast in real (comoving)
space smoothed on a scale $R$ is well described by a lognormal distribution
\citep{coles91},
\begin{equation}
  P(\delta)=\frac{(2 \pi \omega^2_R)^{-1/2}}{1+\delta}
  \exp{ \left(-\frac{(\ln(1+\delta)+\omega^2_R)^2}{2\omega^2_R}\right)
  } \,\,\,\, ,
\label{eq:pdf}
\end{equation} 
where $\omega^2_R=\ln(1+\left<\delta^2\right>_R)$ and
$\left<\delta^2\right>_R=\sigma^2_R(z)$, with $\sigma_R(z)$ being the
standard deviation of mass fluctuations at redshift $z$ on the same scale: 
\begin{equation}
  \sigma_R^2(z)=\int_0^\infty \frac{dk}{k}
  \frac{k^3P(k,z)}{2\pi^2}|W(kR)|^2 \,\,\,\, .
\end{equation}
Here $P(k,z)$ is the mass power spectrum at redshift $z$ in the adopted
cosmology and $W(x)$ is the Fourier transform of the smoothing window
function. For our purpose we use the mass power spectrum of \citet{smith03},
which includes the non-linear evolution of the initial mass fluctuations
field.

The density field recovered from redshift surveys is affected by galaxy
peculiar motions. Therefore one needs to convert the real-space PDF model into
redshift-space in order to be able to compare it to observations. It has been
found that the redshift-space PDF of the density contrast is still well
described by a lognormal distribution \citep{sigad00}, with a standard
deviation $\sigma^z_R(z)$ related to that in real space as \citep{kaiser87},
\begin{equation}
  \sigma^z_R(z)=\left(1+\frac{2}{3}f(z)+\frac{1}{5}f^2(z)
  \right)^\frac{1}{2} \sigma_R(z) \,\,\,\, .
\end{equation}
Here $\sigma_R$ and $\sigma^z_R$ are respectively the real- and redshift-space
standard deviations and $f(z)$ denotes the growth rate of structure, which in
the framework of General Relativity is well approximated as
$f(z)\simeq\Omega^{0.55}_m(z)$ \citep[][]{wang98}.

Following this procedure we obtain the PDF of the mass density contrast in
redshift-space.  To obtain that of galaxies we have to further apply a biasing
factor. For our purposes here, we simply assume linear deterministic biasing,
setting $\delta_g=b\delta$ \citep[however, see][]{marinoni05}.  We choose a
value $b=\sqrt{2.05}$, as required to match the large-scale amplitude of the
two-point correlation function of galaxies in our sample, as we shall show in
Sec. \ref{sec:results}.

By definition, the PDF described by equation \ref{eq:pdf} refers to the
distribution of $\delta$ as measured in randomly placed spheres within the
survey volume.  On the other hand, for the data we have at our disposal only
the {\sl conditional} values of $\delta_g$ as measured in volumes centred on
each galaxy in the sample.  Given the probabilistic meaning of the
distribution function of the density $\mathcal{P}(\rho)$ in the two cases they
must be related as
\begin{equation} 
\mathcal{P}_c(\rho)=\frac{\rho \mathcal{P}(\rho)}{\displaystyle \int_0^\infty \rho
\mathcal{P}(\rho) d\rho}=\frac{\rho}{\bar{\rho}} \mathcal{P}(\rho) \,\,\,\, .
\end{equation}
Being $P(\delta)=\bar{\rho} \mathcal{P}(\rho)$, the corresponding
relation between the PDFs of the density contrast $P(\delta)$ is  
\begin{equation}
P_c(\delta)=\frac{(1+\delta)P(\delta)}{\overline{1+\delta}} \,\,\,\, ,
\end{equation}
where $\overline{1+\delta}=1$.

\begin{figure}
\resizebox{\hsize}{!}{\includegraphics{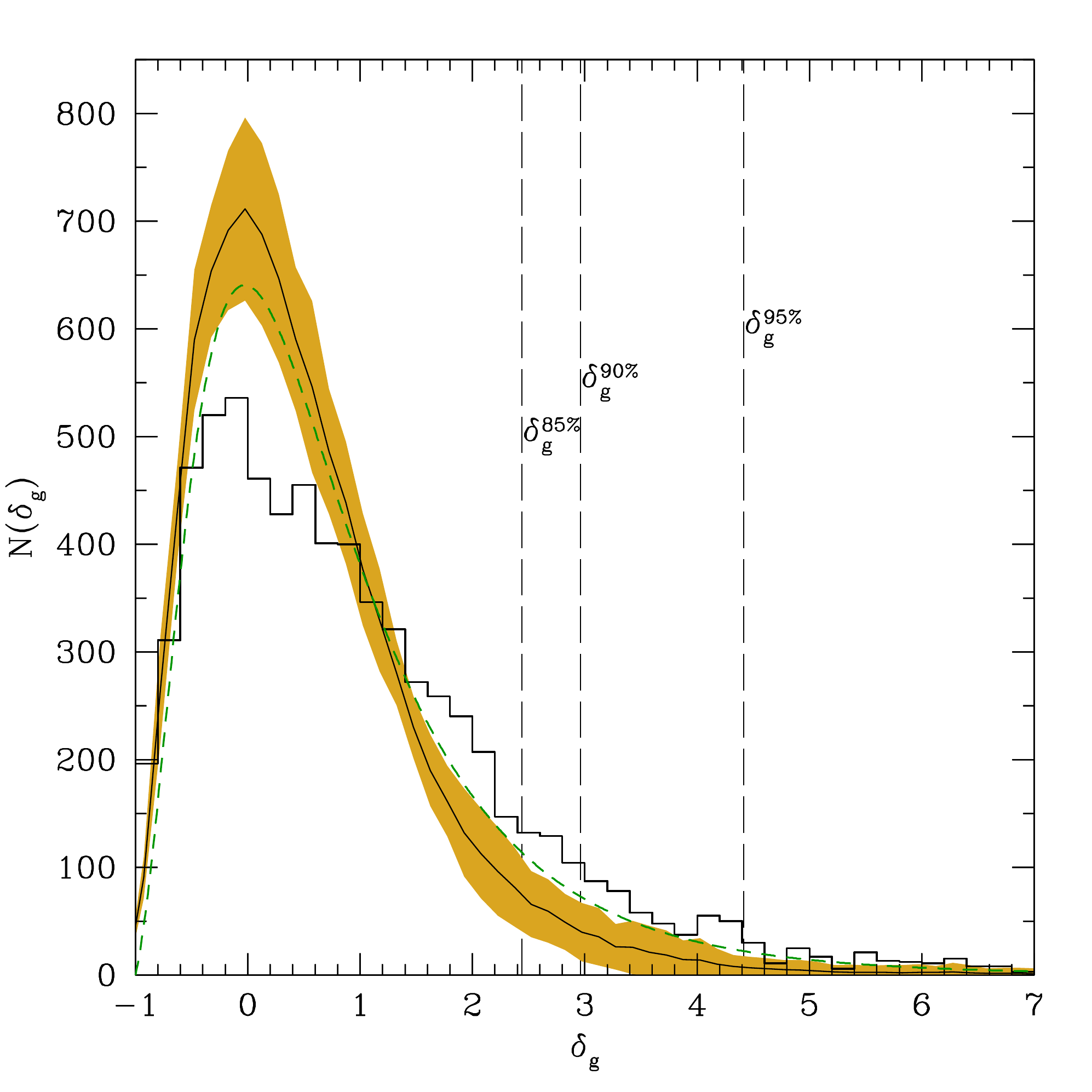}}
\caption{The probability distribution function of the density contrast,
  measured around each galaxy in the current zCOSMOS 10k catalogue as
  discussed in the text (histogram).  The solid line and shaded area
  correspond respectively to the mean and 1$\sigma$ dispersion of the same
  statistics, measured on the 24 Millennium mocks; the dashed curve gives
  instead, as reference, the expected theoretical distribution for a lognormal
  model in a $\Lambda CDM$ cosmology with
  $(\Omega_m,\Omega_\Lambda,\sigma_8)=(0.25,0.75,0.8)$, at the mean redshift
  of the 10k sample, computed as discussed in the text.  The vertical solid
  lines correspond to values of the density contrast excluding the top $5\%$,
  $10\%$ and $15\%$ of the distribution.}
\label{fig:pdf}
\end{figure}

We are then in the position to compare the theoretically predicted $P_c$
(normalised to the total number of galaxies in the sample) to the observed
distribution.  This is presented in Fig.~\ref{fig:pdf}, together with the mean
and scatter (68\% confidence corridor) of the 24 mock samples.  The analytical
prediction and the mean of the mocks are in fair agreement (although they
disagree in the details at the 1$\sigma$ level). Note, however, that the
detailed shape and amplitude of the analytical prediction are quite sensitive
to the choice of the effective redshift $\bar z$ of the survey.  Here we have
used the mean value $\bar z\simeq 0.56$ yielded by the actual redshift
distribution $dN/dz$ of the survey, but using e.g. $\bar z\simeq 0.6$ would
give a better agreement with the PDF from the simulations.  Additionally, the
analytical prediction cannot include the small scale ``Finger-of-God'' effect
due to high velocities in clusters (which however has the effect to reduce
power on small scales). Finally, it has been computed using the more
up-to-date $\sigma_8=0.8$, to check the impact of the value $\sigma_8=0.9$
used for the Millennium Run.  Beyond these points, the simple goal of the
analytical model is to show an alternative -- yet more idealised -- example,
in addition to the mocks, of what one should expect from the theory.  What is
relevant for this work is that the conditional PDF of the data differs
strongly from both theoretical predictions.  Peaking at $\delta_g\simeq -0.2$,
it shows an extended high-density tail out to $\delta_g\simeq 7$.  The
distribution expected from the models is more peaked around $\delta_g=0$ and
drops more rapidly for $\delta_g>2$.  This plot clearly shows a statistically
significant excess of high-density regions in the galaxy data.
\footnote{An ongoing analysis of the fuller COSMOS sample to
  $I_{AB}=24$ based on
  photometric redshifts (Scoville et al. 2010, in preparation), seems
  to indicate
  a better agreement of the observed PDF to that of the Millennium
  mocks.  This might be explained as a consequence of the larger volume of the
  sample used (lower cosmic variance), together with the fairly large
  smoothing window used to define the over-densities and, most 
  importantly, the blurring of the PDF produced by the photometric
  redshift errors.}

\subsection{Clustering estimation}

We estimate real-space galaxy clustering using the standard projected
two-point correlation function, $w_p(r_p)$, that properly corrects for
redshift-space distortions due to galaxy peculiar motions. This is obtained by
projecting the two-dimensional two-point correlation function $\xi(r_p,\pi)$
along the line-of-sight\footnote{In practice, a finite value for the the upper
  integration limit is adopted.  We use $20$\hmpc which recovers the signal
  dispersed by redshift distortions, while minimising the noise that dominates
  at large values of $\pi$ \citep[see also][]{meneux09,porciani09}.}
\begin{equation}
  w_p(r_p)=2\int_{0}^{\infty}\xi(r_p,\pi)d\pi \,\,\,\,  ,\label{eq:wprp}
\end{equation}
where $r_p$ and $\pi$ are the components of the galaxy-galaxy separation
vector respectively perpendicular and parallel to the line-of-sight
\citep{peebles80,fisher94}. $\xi(r_p,\pi)$ is measured using the
\citet{landy93} estimator and properly accounting for the survey selection
function and various incompleteness effects, as thoroughly described in
\citet{delatorre09}.  Error bars are estimated through the blockwise bootstrap
method \citep[e.g.][]{porciani02,norberg09}. This is discussed in detail and
compared to results from mock samples in \citet[][]{meneux09} and
\citet{porciani09}. All clustering codes and methods used here have been
extensively tested against independent programs in the course of the latter
analyses.

\section{Results and Discussion} \label{sec:results}

\begin{figure}
\resizebox{\hsize}{!}{\includegraphics{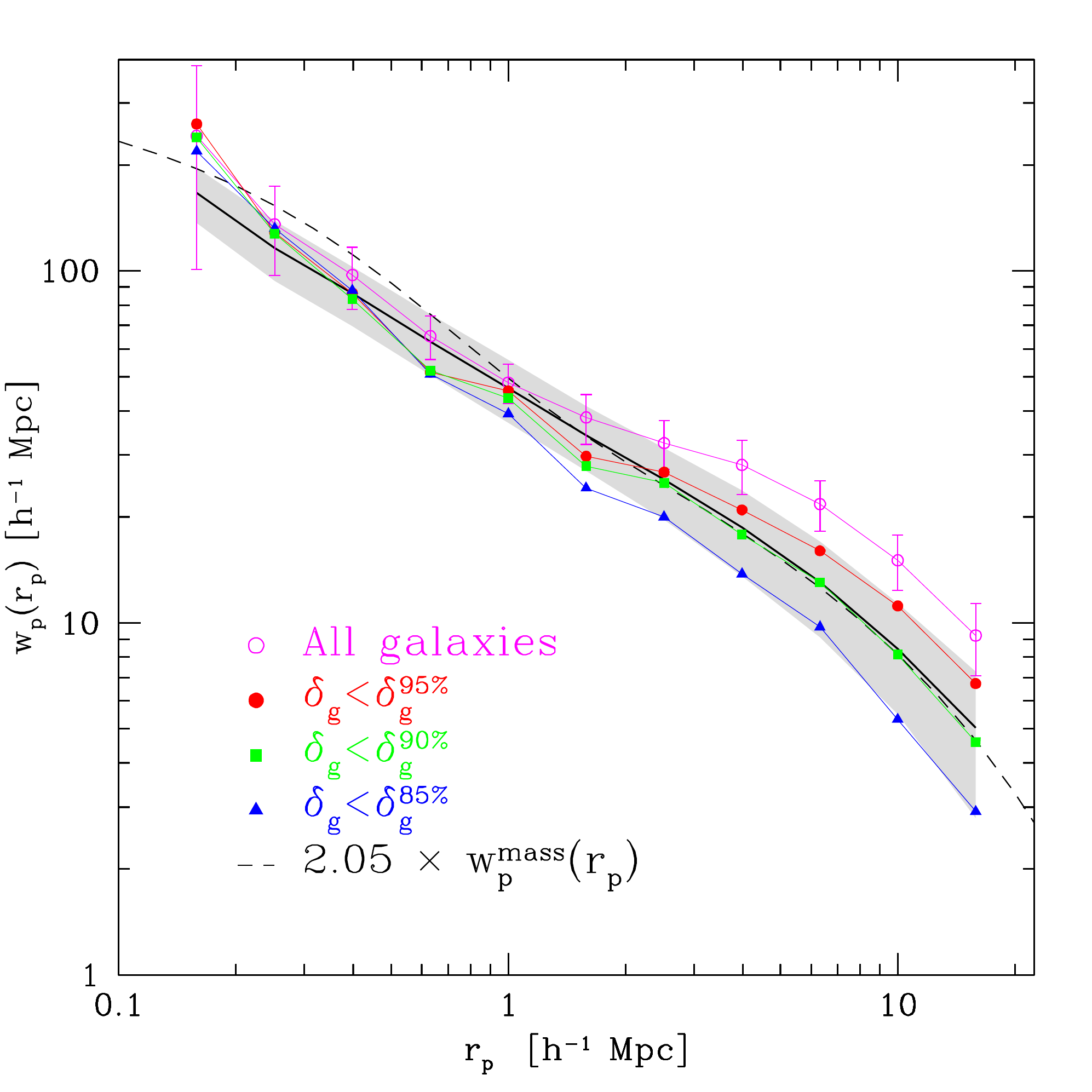}}
\caption{The projected two-point correlation function \wprp of the zCOSMOS 10k
  at $0.6<z<1$, compared to sub-samples in which galaxies living in the
  densest environments are gradually excluded (top to bottom). To reduce
  confusion, error bars are shown for the main sample only, being in general
  of amplitude comparable to the scatter of the mock samples indicated by the
  shaded area. The thick solid line and surrounding shaded corridor correspond
  in fact to the mean and $1\sigma$~scatter of the $24$ mock surveys.  For
  comparison, the dashed curve also shows the \emph{HALOFIT} \citep{smith03}
  analytic prescription for the non-linear mass power spectrum (assuming
  $\Lambda CDM$ with $(\Omega_m,\Omega_\Lambda,\sigma_8)=(0.25,0.75,0.8)$),
  multiplied by an arbitrary linear bias $b^2=2.05$.  The shape of \wprp for
  zCOSMOS galaxies agrees with the models when the 10\% densest environments
  are eliminated.}
\label{fig:xi}
\end{figure}

In Fig. \ref{fig:xi} we show the projected correlation function \wprp computed
for the 10k sample in the redshift range $0.6<z<1$ (top curve), together with
those from a series of sub-samples in which we gradually eliminated galaxies
located in the most dense environments.  We excluded, respectively, the top
$5\%$, $10\%$, and $15\%$ fractions of the distribution of over-densities,
corresponding to the dashed vertical lines in Fig.~\ref{fig:pdf}.  \wprp for
the full 10k sample shows a very flat shape, with significant ``excess
clustering'' above 1\hmpc, as seen in previous analyses of the COSMOS/zCOSMOS
data.  When galaxies in the densest environments are excluded, however, the
large-scale ``shoulder'' gradually disappears.  What we see is a clear
dependence of the mean large-scale clustering of galaxies on the type of
environments they inhabit, similarly to the results of \citet{abbas07} from
the SDSS.

\begin{figure}
\resizebox{\hsize}{!}{\includegraphics{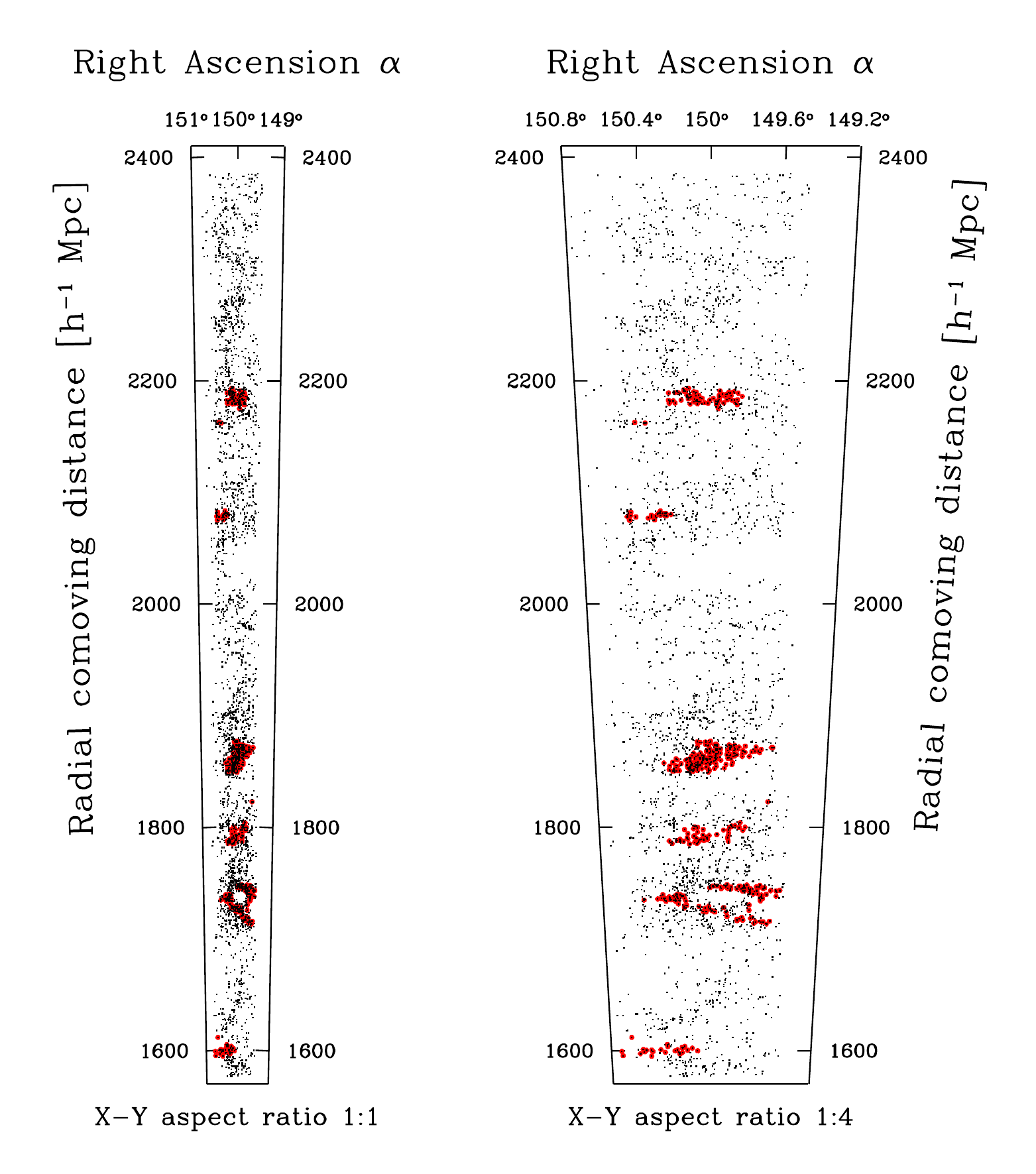}}
\caption{The spatial distribution of galaxies with $0.6<z<1$ (dots) in the
  zCOSMOS 10k sample, highlighting those inhabiting the 10\% highest-density
  tail of the distribution (circles).  These galaxies clearly belong to a few
  well-defined structures. The right-hand plot is simply an expanded version
  of the pencil-beam on the left, to enhance visibility.}
\label{fig:cone}
\end{figure}

In the same figure we also show the ``universal'' \wprp expected in the
standard $\Lambda$CDM cosmological model with linear biasing. The theory
predictions are, again, obtained in two ways.  First, we use \emph{HALOFIT}
\citep{smith03} to compute directly the approximated non-linear mass power
spectrum expected at the survey mean redshift.  Secondly, we compute the
average and scatter of \wprp from the 24 mock samples.  Remarkably, the two
curves (solid and dashed black lines) are virtually indistinguishable above 1
\hmpc once the \emph{HALOFIT} mass correlation function is properly multiplied
by an arbitrary linear bias factor of $b^2=2.05$.  The comparison to the data
shows a very good agreement for the 10k sub-sample in which the 10\% densest
environments were excluded.  We note that the shape of \wprp measured from the
independent VVDS survey shows a shape which is closer to the model predictions
\citep{meneux09}.  With this thresholding in density of the 10k data,
therefore, we are able to bring the measured shape of galaxy clustering at
$z\simeq0.7$ from zCOSMOS, VVDS and the standard cosmological model within
close agreement, suggesting a more quantitative interpretation of the flat
shape of \wprp observed in zCOSMOS at these redshifts.  In previous papers
\citep[e.g.][]{meneux09,kovac09} we already suggested that this could be due
to the presence of particularly significant large-scale structures between
$z=0.5$ and $z=1$.  Here we see that it is in fact driven by an excess of
galaxies sampling high-density regions, skewing the density distribution away
from the supposedly ``Universal''. Fig.~\ref{fig:cone} shows where these
high-density galaxies are actually located within the 10k sample.  The
galaxies belonging to the $10\%$ high-density tail are marked by (red) circles
and turn out to belong to a few very well-defined structures only.\footnote{We
  also directly tested whether the two-point correlation functions computed on
  the density-thresholded samples were by any means sensitive to the way the
  ``depleted'' sub-volumes were treated (e.g. kept in or excluded when
  building the standard reference random sample); no significant changes were
  found.}  It is easy to imagine that if embedded in a larger volume, these
structures would not weight so much as to modify significantly the overall
shape of the PDF.  As seen from the histogram in Fig.~\ref{fig:pdf}, however,
in this volume there is a clear over-abundance of high-density galaxy
environments, while regions with average density are under-represented.

One may wonder however, whether the theoretical model represented by the
Millennium mocks can be taken as a reliable reference.  In fact, it has been
shown that this specific model tends to overestimate the overall amplitude of
\wprp at $z\simeq 1$ and does not reproduce the observed clustering
segregation in colour (e.g. \citealt{coil08}; de la Torre et al., in
preparation) In general, semi-analytical recipes do tend to affect the
amplitude and shape of the correlation function. This however happens only on
small scales, where the complex interplay between galaxy formation processes
and the distribution of dark-matter halos has an impact.  On large scales
instead, they predict a fairly linear biasing, as we can see directly
comparing the solid and dashed black lines in Fig. \ref{fig:xi}.  This means
that the large-scale shape of the correlation function is essentially driven
by the underlying mass distribution in the assumed cosmological model and not
by the details of the semi-analytic recipe adopted to generate galaxies.  A
different recipe would not affect, therefore, the results obtained here,
unless we postulate the existence of dramatically non-local galaxy formation
processes \citep[e.g.][]{narayanan00}.

We also note from Fig. \ref{fig:xi} that the dependence of \wprp on the PDF
threshold is essentially on large scales. Below $\sim 1$\hmpc there is no
significant change when denser and denser environments are excluded.  In their
analysis of the SDSS \citet{abbas07} consider sub-samples defined as extrema
of the density distribution, i.e. using galaxies lying on the tails of the
distribution on both sides. With this selection, they find a change in \wprp
for different environments also on small scales.  It can be shown simply using
the conservation of galaxy pairs \citep[see Eq. 1 of ][]{abbas07} that the two
results are in fact consistent with each other (Ravi Sheth, private
communication).

These results highlight the importance in redshift surveys of an accurate
reconstruction of the density field, to evidence possible peculiarities in the
overall PDF as sampled by that specific catalogue. Further strengthening the
results obtained by \citet{abbas07} at $z\simeq 0$, we have shown that an
anomalous density distribution function can significantly bias the recovered
two-point correlation function, making it difficult to draw general
conclusions from its shape. This result provides another example of the
intrinsic difficulty existing when comparing observations of the galaxy
distribution to theoretical predictions. The theory provides us with fairly
accurate forecasts for the distribution of the dark matter and for that of the
halos within which we believe galaxies form \citep[e.g.][]{mo96,sheth99}.
However, translating galaxy clustering measurements into constraints for the
halo clustering involves understanding how the selected galaxies populate
halos with different mass.  The result presented here show how a sample
particularly rich in dense structures favours higher-mass halos, which in turn
are more clustered, thus biasing the observed correlation function as a
function of scale.  A more detailed analysis of the environmental dependence
of galaxy clustering in the zCOSMOS-Bright sample and related HOD modelling
will be presented in a future paper.

\section*{Acknowledgments}

We thank Ravi Sheth for helpful comments on the manuscript. Financial support
from INAF and ASI through grants PRIN-INAF--2007 and ASI/COFIS/WP3110
I/026/07/0 is gratefully acknowledged. LG thanks D. Sanders and the University
of Hawaii, for hospitality at the Institute for Astronomy, where this work
was initiated.

This work is based on observations undertaken at the European Southern
Observatory (ESO) Very Large Telescope (VLT) under Large Program
175.A-0839. Also based on observations with the NASA/ESA Hubble Space
Telescope, obtained at the Space Telescope Science Institute, operated by the
Association of Universities for Research in Astronomy, Inc. (AURA), under NASA
contract NAS 5Y26555, with the Subaru Telescope, operated by the National
Astronomical Observatory of Japan, with the telescopes of the National Optical
Astronomy Observatory, operated by the Association of Universities for
Research in Astronomy, Inc. (AURA), under cooperative agreement with the
National Science Foundation, and with the Canada-France- Hawaii Telescope,
operated by the National Research Council of Canada, the Centre National de la
Recherche Scientifique de France, and the University of Hawaii.

\bibliography{biblio}

\bsp

\label{lastpage}

\end{document}